\documentclass[useAMS,usenatbib,usegraphicx,referee]{biom}
\pdfoutput=1
\usepackage{amsmath, amsfonts, caption, multirow, tikz} %, pdfpages}

\title[SECR model to estimate trap detection rate]{A spatially explicit capture recapture model for partially identified individuals when trap detection rate is less than one}

\author{Soumen Dey$^{1,*}$\email{soumendey@isibang.ac.in}, 
Mohan Delampady$^{1,**}$\email{mohan@isibang.ac.in},
$\mbox{\bf K. Ullas Karanth}^{2,3,4,***}$\email{ukaranth@gmail.com}, 
$\mbox{\bf Arjun M. Gopalaswamy}^{1,5,****}$\email{arjungswamy@gmail.com}\\
$^{1}$ Statistics and Mathematics Unit, Indian Statistical
Institute, Bangalore Centre, Bengaluru 560059, India\\
$^{2}$ Centre for Wildlife Studies and 
$^{3}$ Wildlife Conservation Society, India program,\\551, 7th Main Road, Rajiv Gandhi Nagar-2nd Phase, Kodigehalli, Bengaluru (Bangalore), Karnataka-560097, India\\
$^{4}$ National Centre for Biological Sciences, Tata Institute of Fundamental \\
Research, Bellary Road, Bengaluru 560065, India\\
$^{5}$ Department of Zoology, University of Oxford, South Parks Road, Oxford OX1 3PS, UK}

\begin{document}

\pagerange{\pageref{firstpage}--\pageref{lastpage}}
\label{firstpage}
\begin{abstract}

Spatially explicit capture recapture (SECR) models have gained enormous popularity to solve abundance estimation problems in ecology. In this study, we develop a novel Bayesian SECR model that disentangles the process of animal movement through a detector from the process of recording data by a detector in the face of imperfect detection. We integrate this complexity into an advanced version of a recent SECR model involving partially identified individuals \citep{royle2015spatial}. We assess the performance of our model over a range of realistic simulation scenarios and demonstrate that estimates of population size $N$ improve when we utilize the proposed model relative to the model that does not explicitly estimate trap detection probability \citep{royle2015spatial}. We confront and investigate the proposed model with a spatial capture-recapture data set from a camera trapping survey on tigers (\textit{Panthera tigris}) in Nagarahole, southern India. Trap detection probability is estimated at 0.489 and therefore justifies the necessity to utilize our model in field situations. We discuss possible extensions, future work and relevance of our model to other statistical applications beyond ecology. 
\end{abstract}
\maketitle
\section{Introduction}\label{intro}
Understanding the dynamics of wildlife populations is central to answering ecological questions and forms the basis for conservation. However, owing to sampling problems (primarily imperfect detection and spatial sampling) \citep{williams2002analysis},
%(\citealt{yoccoz2002age})
it is a major challenge to accurately characterise wildlife populations from field data to reliably estimate state variable parameters. The challenge is greater when the species is cryptic, occurs at low density and often elusive, as with large carnivores \citep{karanth1995estimating}
%, jackson2006estimating}
 and rare ungulates \citep{obrien2011estimation}. This problem has motivated the development of several tailor-made statistical estimators over the years (\citealt{williams2002analysis}, \citealt{buckland2001introduction}, \citealt{mackenzie2006occupancy}).
%  \citealt{royle2014spatial}). 

More recently, such classes of ecological problems have been addressed elegantly using hierarchical models, where a distinction between the `state process' (the true state of the ecological system that is of main interest) and the `observation process' (the way in which observations occur during sampling) are explicitly defined in the modelling (\citealt{royledorazio2008hierarchical}, \citealt{banerjee2014hierarchical}). 
%\citealt{kery2015applied}).
  Based on this philosophy, the development of spatially explicit capture-recapture models (hereafter SECR models) (\citealt{borchers2008spatially}, \citealt{royle2009bayesian}) for estimating animal abundance has witnessed an explosive growth  \citep{royle2014spatial}. Under this approach, observation data about individuals are recorded by spatial array of detectors (such as camera traps, hair snares, fixed traps) within an area of interest over a fixed time period. SECR models utilize the spatial locations of animal `detections' to explicitly enable inference about the spatial distribution of animals in addition to estimating animal abundance and has seen wide application for globally threatened species (\citealt{royle2009bayesian}, \citealt{broekhuis2016counting}, \citealt{elliot2017toward} for examples related to \textit{Panthera tigris} (tigers), \textit{Acinonyx jubatus} (cheetahs) and \textit{Panthera leo} (lions) respectively) to better understand animal clustering processes or to identify target regions for conservation. 

However, all these inferences are drawn from data emanating when animals pass through a spatial array of detectors. Currently, SECR models do not account for the fact that detectors often perform imperfectly. Furthermore local, detector-level, effects may explain whether an animal will pass through the detector or not. For example, workers often use baits to attract animals to trap stations when animals are in the vicinity and investigators are often interested to understand animal response to such detectors. However, an animal passing through a detector will not necessarily mean that the detector will record this event perfectly. A failure to recognize this distinction will mislead inferences about the ecological process of animal movement and distribution. If we assume that over a fixed number of detection attempts at a location we will detect an animal with certainty, then a newer development \citep{clare2017pairing} can be used to address this problem. However, this is a restrictive assumption to meet in the real world.   
%\subsection{Study objectives} 
\paragraph{Study objectives}
Thus, in this study: (1) We develop an SECR estimator that disentangles the process of animal movement through a detector from the process of device performance by utilizing information on event captures on multiple devices at particular locations. Further, we integrate this complexity into an advanced version of the SECR model involving partial identifications of individuals (\citealt{royle2015spatial}, \citealt{augustine2016spatial}). 
(2) We assess the performance of our model over a range of realistic simulation scenarios typically faced in field ecological studies of large, charismatic, wildlife species. 
(3) We confront this model with a spatial capture-recapture data set from a long term camera trapping survey on tigers (\citealt{karanth2017spatiotemporal}, \citealt{dorazio2017hierarchical}) 
%(CITE KARANTH ET AL, PROC SOC PAPER and DORAZIO AND KARANTH 2017).  
(4) We discuss possible extensions, future work and relevance of our model to other statistical applications beyond ecology. 
\section{Methods}\label{methods}
In typical photographic capture-recapture surveys (\citealt{royle2009bayesian}, \citealt{Oconnell2011camera})
% \citep[see][]{royle2009bayesian, Oconnell2011camera}
an array consisting of camera trap stations is placed to sample a species of interest. Each station comprises of two cameras (detectors) meant to capture both flank images of animals. If the species is naturally marked individuals are identified by their unique markings. Often, however, the detectors perform imperfectly, leading to single flank images leading to problems of reconciliation of individual identities. While, this example motivated our specific model development, we can also envision many scenarios where more than one detector is used to extract features of individual identity of an animal at each station. 
%But identifying individuals from field data is not straightforward and often individual detection histories are obtained by multiple detectors which may not be reconcilable. This, in turn, poses challenge to model recorded data. We shall talk about this issue synchronisation in later sections. First let us give a short description on the sampling design. 
%Let us primarily focus on probabilistic disentanglement of two key parameters.
% By ignoring the need of synchronisation while modelling the data will result in overestimation of the abundance and underestimation of standard errors and below nominal coverage for credible interval estimates, as pointed out by \cite{bonner2013mark}.  Statistical models have been developed to address the problem of misidentification or partial identification along with imperfect detection process in the above mentioned capture recapture studies - \cite{royle2015spatial}, \cite{augustine2016spatial}, \cite{mcclintock2013integrated}, \cite{bonner2013mark} (for sampling via camera trapping), \cite{wright2009incorporating} (for sampling via collection of genetic material).
\subsection{Modelling approach} \label{crdata}
We utilize the hierarchical modelling philosophy \citep{royledorazio2008hierarchical}
%\citealt{kery2015applied})
to formulate a model to incorporate the problem of imperfect detection of detectors in spatial capture-recapture models. A list of notations used in this article is provided in Tables~\ref{par.definitons1} and \ref{par.definitons2}. 
\subsubsection{State process}\label{stateprocess}
Consider a population of individuals of certain species that reside within a bounded, geographic region $\mathcal{V} \, (\subset \mathbb{R}^2)$ that has scientific or operational relevance. Each individual is assumed to be located following a point process \citep{borchers2008spatially} by having an activity centre located at $\mathbf{s} \, (\in \mathcal{V}$). 
%However, we do not know how many animals ($N$) there are and it is our goal to estimate this parameter. 
Let $\mathbf{S}$ denote an array of latent variables defining the locations of the $N$ (unknown) animals in the study. For the ease of computation and other technical advantages (described later), we define $N \sim \mathrm{Binomial}\left(M,\psi\right)$, where $M$ represents the maximum possible number of individuals present within $\mathcal{V}$ and $\psi$ is a thinning parameter to indicate the proportion of $M$ that represent the real population. It should be noted here, that the $N$ animals located at $\mathbf{S}$ are assumed to move around $\mathbf{S}$ according to some prescribed density kernel during the period of sampling. However, previous SECR models regard this movement inherently as part of the observation process (\citealt{borchers2008spatially}, \citealt{royle2009bayesian}, \citealt{royle2015spatial}).%   
\subsubsection{Observation process}\label{observationprocess}
We suppose that a spatial array of $K$ trap stations are placed in the state space $\mathcal{V}$. We consider the situation that two detectors are deployed at each of these trap stations and are kept active for  $J$ sampling occasions. We assume that each detector captures some mutually exclusive attribute of an individual. For example, in a camera trapping survey with two co-located detectors placed at each trap station, detector 1 may represent captures of left flank images of naturally marked animals, such as tigers or leopards, and detector 2 may represent right flank captures. Let $y_{ikt}^{(1)}$ and $y_{ikt}^{(2)}$ represent the Bernoulli capture outcomes for an individual $i$ at trap station $\bf u_k$ on sampling occasion $t$ for detectors 1 and 2 respectively. We note here that it is possible to ascertain the identity of an individual without a doubt only when \emph{both} the detectors record the individual, simultaneously, on at least one occasion during the survey. 
We will suppose that at the end of this survey $n$ individuals are captured and fully identified. Thus, the recorded observations obtained by detectors 1 and 2 are individual-specific detection histories, $\mathbf{Y_{obs}^{(1)}} = ((y_{ikt}^{(1)}))$, $\mathbf{Y_{obs}^{(2)}} = ((y_{ikt}^{(2)}))$, respectively. This implies that the rows of $\mathbf{Y_{obs}^{(1)}}$ and $\mathbf{Y_{obs}^{(2)}}$ are of the same order and of same length $n$ and the dimension of each of the two arrays (of detectors 1 and 2) is $n\times K \times J$. Further, the paired Bernoulli outcomes $y_{ikt} = (y_{ikt}^{(1)}, y_{ikt}^{(2)})$ give rise to bilateral capture-recapture data for each individual $i$ at location $\mathbf{u_k}$ on occasion $t$. So, for an individual $i$, we denote the bilateral capture history by $\mathbf{Y_{i, obs}} = (\mathbf{Y^{(1)}_{i, obs}}, \mathbf{Y^{(2)}_{i, obs}}) = ((y_{ikt}^{(1)}, y_{ikt}^{(2)}))_{k,t}$, which is of dimension $2\times K \times J$. %
\begin{table}[!htbp]
\centering
\caption{\textit{Notations of variables, parameters and latent variables which are used in this article. Note that bold symbols represent collections (vectors) of parameters.}}
\resizebox{0.85\textwidth}{0.48\textheight}{\begin{tabular}{ l @{\extracolsep{15pt}} l } 
\\[-1.8ex]\hline 
\hline \\[-1.8ex]
\textbf{Variables and parameters} & \textbf{Definition}\\ 
\hline \\[-1.8ex]
$\mathcal{V}$ & A bounded geographic region of scientific or operational relevance\\
& where a population of individuals of certain species reside. \\ [+1.5ex]
$N \sim \mathrm{Binomial}(M,\psi)$ & Population size of the superpopulation, i.e., the number of individuals \\  & within $\mathcal{V}$.  \\ [+1.5ex]
$\mathbf{S}$ & Locations of the activity centres of $N$ animals within $\mathcal{V}$.\\ [+1.5ex]
$M$ & Maximum number of individuals within the state space $\mathcal{V}$. \\ & This is a fixed quantity defined by the investigator. \\ [+1.5ex]
$\psi$ & Proportion of individuals that are real and present within $\mathcal{V}$.\\ [+1.5ex]
$K$ & Number of trap stations in $\mathcal{V}$.\\ [+1.5ex]
$J$ & Number of sampling occasions.\\ [+1.5ex]
$p_0$ & Baseline trap entry probability, i.e., probability that an individual \\
& passes through a trap station assuming its centre of activity \\
& is also located at that trap station.\\ [+1.5ex]
$\sigma$ & $\sigma$ measures the spatial extent of movement around individual activity\\  
&  centre. $\sigma = \sigma_m$ for male individuals,  $\sigma = \sigma_f$ for female individuals.\\ [+1.5ex]
% $\sigma_f$ & Spatial extent of movement around individual activity centre\\ 
% & for a female individual. This cannot exceed $D$.\\ [+1.5ex]
$d_{ik} = d(\mathbf{s_i}, \mathbf{u_k}) =  \Arrowvert \mathbf{s_i} - \mathbf{u_k} \Arrowvert$ & Euclidean distance between points $\mathbf{s_i}$ and $\mathbf{u_k}$.\\ [+1.5ex]
$\pi_{ik} = p_0\, \exp(-\frac{d_{ik}^2}{2\sigma^2})$ & Probability that an individual $i$ is present at a trap $k$ at \\
& some occasion $t$. This is a derived parameter. \\ [+1.5ex]
$\theta$ & Probability that an individual is male.\\  [+1.5ex]
$\mathbf{p^{(12)}} = (p^{(12)}_1, p^{(12)}_2, \dots, p^{(12)}_K)$ & $p^{(12)}_k$ denotes the probability of simultaneous detection by both\\
&  detectors 1 and 2 at trap station $\mathbf{u_k}$ during a sampling occasion.\\ [+1.5ex]
$\mathbf{p^{(1\overline{2})}} = (p^{(1\overline{2})}_1, p^{(1\overline{2})}_2, \dots, p^{(1\overline{2})}_K)$ & $p^{(1\overline{2})}_k$ denotes the probability of detection only by detector 1 at a trap\\ 
&  station $\mathbf{u_k}$ during a sampling occasion.  \\ [+1.5ex]
$\mathbf{p^{(\overline{1}2)}} = (p^{(\overline{1}2)}_1, p^{(\overline{1}2)}_2, \dots, p^{(\overline{1}2)}_K)$ & $p^{(\overline{1}2)}_k$ denotes the probability of detection only by detector 2 at trap \\
&  station $\mathbf{u_k}$ during a sampling occasion. \\ [+1.5ex]
$\mathbf{p} = (\mathbf{p^{(12)}},\, \mathbf{p^{(1\overline{2})}},\, \mathbf{p^{(\overline{1}2)}})$ & Collection of all detection probabilities.\\ [+1.5ex]
$\phi_k$ & Probability that an individual $i$ is detected by detector $k$ at a certain \\ 
& occasion, given that it is present at that trap station, $k = 1,2,\dots,K$.\\ [+1.5ex]
$\phi$ & Probability that an individual $i$ is detected by a detector on \\
& some occasion $t$ given that it is present at that trap.\\ [+1.5ex]
$R$ & Maximum permissible value of movement range for each individual\\
& during the study.\\
[+3ex]
\textbf{Latent variables} & \textbf{Definition}\\ 
\hline \\[-1.8ex]
$\mathbf{s_i} = (s_{i1}, s_{i2})^\prime$ & Location of individual $i$'s activity centre. \\ [+1.5ex]
$E_{ikt}$ & $E_{ikt} = 1$ if individual $i$ has entered a trap station $k$ on occasion $t$,\\
& $E_{ikt}=0$ if not entered.\\ [+1.5ex]
$\mathbf{z}=(z_1, z_2, \dots, z_M)^\prime$ & A vector of Bernoulli variables, $z_i=1$ if individual $i$ is present. \\ [+1.5ex]
$\mathbf{x}=(x_1,  \dots, x_M)^\prime$ & A vector of Bernoulli variables, $x_i=1$ if individual $i$ is male \\ 
& in the population and $x_i=0$ if it is a female.\\ [+1.5ex]
$\mathbf{x_0}\ (\subset \mathbf{x})$ & Vector of `missing' binary observations on genders of the\\
& list of $M$ individuals.\\ [+1.5ex]
%\multirow{3}{*}{$\mathbf{L}=(\text{L}_1, \text{L}_2, \dots, \text{L}_M)^\prime$} 
$\mathbf{L}=(\text{L}_1, \text{L}_2, \dots, \text{L}_M)^\prime$ & $\mathbf{L}$ is a one to one mapping from index set of individuals from \\
&  detector 2 to $\{1,2,\dots,M\}$ providing the true index of each\\
&  detector 2 individuals.\\ [+3ex]
%$E_{ik0}=\displaystyle{\sum_{t=1}^J E_{ikt}} $ & Number of times individual $i$  is present at trap $k$ \\
%& over $J$ occasions.\\
%$E_{i00}=\displaystyle{\sum_{k=1}^K \sum_{t=1}^J E_{ikt}} $ & Number of times individual $i$ is present at trap $k$  \\
%& over $K$ traps and $J$ occasions.\\[+3ex]
\hline \\[-1.8ex] 
\end{tabular}}
 \label{par.definitons1}
\end{table}
\begin{table}[!htbp]
\centering
\caption{\textit{Notations of data which are used in this article. Note that bold symbols represent collections (vectors) of parameters.}}
%\footnotesize{
\footnotesize{\begin{tabular}{ l @{\extracolsep{15pt}} l } 
\\[-1.8ex]\hline 
\hline \\[-1.8ex]
\textbf{Data } & \textbf{Definition}\\ 
\hline \\[-1.8ex]
 $\mathbf{u_k}= (u_{k1}, u_{k2})^\prime$ & $k^{\,\text{th}}$ trap station for detectors.\\  [+1.5ex]
$y_{ikt}^{(1)}$ & $y_{ikt}^{(1)}=1$ if individual $i$ is detected in detector 1 at trap station \\
& $\mathbf{u_k}$ on occasion $t$, $y_{ikt}^{(1)}=0$ if not detected in detector 1.\\
%$y_{ik0}^{(1)}=\displaystyle{\sum_{t=1}^J y_{ikt}^{(1)}} $ & Number of times individual $i$ got detected in detector 1 \\
%& at trap station $\mathbf{u_k}$ over $J$ occasions.\\
$y_{i\cdot \cdot}^{(1)}=\displaystyle{\sum_{k=1}^K \sum_{t=1}^J y_{ikt}^{(1)}} $ & Number of times individual $i$ got detected\\ [-1.8ex]
&  in detector 1 over $K$ trap stations and $J$ occasions.\\ [+1.5ex]
$y_{ikt}^{(2)}$ & $y_{ikt}^{(2)}=1$ if individual $i$ is detected in detector 2 at trap station  \\
& $\mathbf{u_k}$ on occasion $t$, $y_{ikt}^{(2)}=0$ if not detected in detector 2.\\
%$y_{ik0}^{(2)}=\displaystyle{\sum_{t=1}^J y_{ikt}^{(2)}} $ & Number of times individual $i$ got detected in detector 2 \\
%& at trap station $\mathbf{u_k}$ over $J$ occasions.\\
$y_{i\cdot \cdot}^{(2)}=\displaystyle{\sum_{k=1}^K \sum_{t=1}^J y_{ikt}^{(2)}} $ & Number of times individual $i$ got detected\\ [-1.8ex]
&  in detector 2 over $K$ trap stations and $J$ occasions.\\ [+1.5ex]
$n$  & Number of fully identified individuals, each of them is captured\\
 & by both the detectors on at least one occasion.\\ [+1.5ex]
 $\mathbf{Y_{obs}^{(1)}} = ((y_{ikt}^{(1)}))_{i,k,t}$ & Array of individual specific capture histories  obtained by\\
 & detector 1 (dimension $n \times K \times J$).\\ [+1.5ex]
 $\mathbf{Y_{obs}^{(2)}} = ((y_{ikt}^{(2)}))_{i,k,t}$ & Array of individual specific capture histories  obtained by\\
 & detector 2 (dimension $n \times K \times J$).\\ [+1.5ex] 
 $\mathbf{Y^{(1)}}$ & Zero augmented array of individual specific capture histories \\
 & corresponding to detector 1 (dimension $M \times K \times J$).\\ [+1.5ex]
 $\mathbf{Y^{(2)}}$ & Zero augmented array of individual specific capture histories \\
 & corresponding to detector 2 (dimension $M \times K \times J$).\\ [+1.5ex]
% $\mathbf{Y} = (\mathbf{Y^{(1)}}, \mathbf{Y^{(2)}}) = ((y_{ikt}^{(1)}, y_{ikt}^{(2)}))_{i,k,t}$ &   Combined array of zero augmented capture histories.\\ 
% & (dimension $2\times M \times K \times J$). \\ [+1.5ex]
 $\mathbf{x_{obs}} \ (\subset \mathbf{x})$ & Vector of `recorded' binary observations on genders of the\\
 & captured individuals.\\ [+1.5ex]
 $\mathbf{Y^{(2*)}}$ & Reordered $\mathbf{Y^{(2)}}$ according to $\mathbf{L}$ (dimension $M \times K \times J$).\\ [+1.5ex]
 $n_{ik}=\displaystyle{\sum_{t=1}^J} \,I(y_{ikt}^{(1)} + y_{ikt}^{(2)} >0)$ & Number of times individual $i$ got detected at trap $k$ \\[-1.8ex]
 & on at least one of its sides over $J$ occasions. \\ 
 $n_{i\cdot}=\displaystyle{\sum_{k=1}^K} n_{ik}$ & Number of times individual $i$ got detected on at least\\[-1.8ex]
 & one of its sides over $K$ traps and $J$ occasions. \\ [+1.5ex]
\hline \\[-1.8ex] 
\end{tabular}}
 \label{par.definitons2}
\end{table}
\begin{example} \label{example:sampledata}
In a survey, consider paired detectors (1 and 2), deployed at each of 3 ($ = K$) trap stations and active for 4 ($=J$) sampling occasions. From this survey, we suppose that 2 ($=n$) distinct individuals were \emph{fully identified} since we obtained at least one simultaneous capture (caught at the same time in both detectors) during the survey. Detection histories are thus presented in Table~\ref{sampledata}. For each of the fully identified individuals, the dimension of the detection history data set is $2 \times 3 \times 4$.
\end{example}
\begin{table}[!htbp]
 \centering 
 \caption{\textit{An example of detection histories generated from \ref{example:sampledata}. The survey yield 2 fully-identified individuals and detection histories of partially-identified individuals. The circled outcome corresponds to the detection event that assists in the reconciliation of an individual identity because the two detectors simultaneously captured the animal on this particular sample point. For example, individual 1 was fully-identified owing to the capture event on trap 1 on occasion 2. Due to mutual exclusivity of capture events in the detection histories of the partially-identified individuals, we are uncertain about whether these histories correspond to two different individuals or to the same individual.}} 
{\footnotesize \begin{tabular}{l @{\extracolsep{15pt}} c @{\extracolsep{15pt}} c @{\extracolsep{15pt}} c @{\extracolsep{15pt}} c @{\extracolsep{15pt}} c @{\extracolsep{30pt}} c @{\extracolsep{15pt}} c @{\extracolsep{15pt}} c @{\extracolsep{15pt}} c @{\extracolsep{15pt}} c} 
%{\small \begin{tabular}{l{2cm}l{2cm}c{2cm}c{2cm}c{2cm}c{2cm}l{2cm}c{2cm}c{2cm}c{2cm}c} 
\\[-1.8ex]\hline 
\hline \\[-1.8ex] 
& & \multicolumn{4}{c}{Detector 1} & & \multicolumn{4}{c}{Detector 2}\\
& Occasion & 1 & 2 & 3 & 4 & Occasion & 1 & 2 & 3 & 4 \\
& Trap & & & & & Trap & & & &   \\ 
\hline \\[-1.8ex] 
\multirow{3}{*}{Fully-identified individual 1} 
& 1 & 0 & \raisebox{0.5pt}{\textcircled{\raisebox{-0.9pt} {1}}} & 0 & 0 & 1 & 1 & \raisebox{0.5pt}{\textcircled{\raisebox{-0.9pt} {1}}} & 0 & 0  \\ 
& 2 & 0 & 0 & 0 & 0 & 2 & 0 & 1 & 1 & 0  \\
& 3 & 1 & 0 & 0 & 0 & 3 & 0 & 1 & 0 & 0  \\  
\hline \\[-1.8ex] 
\multirow{3}{*}{Fully-identified individual 2} 
& 1 & 0 & 0 & 0 & \raisebox{0.5pt}{\textcircled{\raisebox{-0.9pt} {1}}} & 1 & 1 & 1 & 0 &  \raisebox{0.5pt}{\textcircled{\raisebox{-0.9pt} {1}}}  \\ 
& 2 & 1 & 0 & 0 & 1 & 2 & 0 & 0 & 0 & 0  \\
& 3 & 0 & 1 & 0 & 0 & 3 & 1 & 0 & 1 & 0  \\
\\[-1.8ex]\hline 
\hline \\[-1.8ex]
\multirow{3}{*}{Partially-identified individual} 
& 1 & 0 & 1 & 0 & 1 & 1 & - & - & - & -  \\ 
& 2 & 1 & 0 & 1 & 0 & 2 & - & - & - & -  \\
& 3 & 0 & 0 & 0 & 1 & 3 & - & - & - & - \\  
\hline \\[-1.8ex] 
\multirow{3}{*}{Partially-identified individual} 
& 1 &- & - & - & - & 1 & 1 & 0 & 1 & 0  \\ 
& 2 & - & - & - & - & 2 & 0 & 1 & 0 & 0  \\
& 3 & - & - & - & - & 3 & 1 & 0 & 1 & 0  \\  
\hline \\[-1.8ex] 
\end{tabular}}
 \label{sampledata} 
\end{table}
%\begin{table}[!htbp]
% \centering 
% \caption{\textit{An example of detection history from detector 1 and 2 with 2 detected individual, 3 traps and 4 detected individuals. The rows index individual, even across different detectors.}} 
%{\footnotesize \begin{tabular}{l @{\extracolsep{15pt}} c @{\extracolsep{15pt}} c @{\extracolsep{15pt}} c @{\extracolsep{15pt}} c @{\extracolsep{15pt}} c @{\extracolsep{30pt}} c @{\extracolsep{15pt}} c @{\extracolsep{15pt}} c @{\extracolsep{15pt}} c @{\extracolsep{15pt}} c} 
%%{\small \begin{tabular}{l{2cm}l{2cm}c{2cm}c{2cm}c{2cm}c{2cm}l{2cm}c{2cm}c{2cm}c{2cm}c} 
%\\[-1.8ex]\hline 
%\hline \\[-1.8ex] 
%& & \multicolumn{4}{c}{Detector 1} & & \multicolumn{4}{c}{Detector 2}\\
%& Occasion & 1 & 2 & 3 & 4 & Occasion & 1 & 2 & 3 & 4 \\
%& Trap & & & & & Trap & & & &   \\ 
%\hline \\[-1.8ex] 
%\multirow{3}{*}{Individual 1} 
%& 1 & 0 & 1 & 0 & 0 & 1 & 1 & 1 & 0 & 0  \\ 
%& 2 & 0 & 0 & 0 & 0 & 2 & 0 & 1 & 1 & 0  \\
%& 3 & 1 & 1 & 0 & 0 & 3 & 0 & 1 & 0 & 0  \\  
%\hline \\[-1.8ex] 
%\multirow{3}{*}{Individual 2} 
%& 1 & 0 & 0 & 0 & 1 & 1 & 1 & 1 & 0 & 1  \\ 
%& 2 & 1 & 0 & 0 & 1 & 2 & 0 & 0 & 0 & 0  \\
%& 3 & 0 & 1 & 0 & 0 & 3 & 0 & 1 & 1 & 0  \\  
%\hline \\[-1.8ex] 
%\end{tabular}}
% \label{sampledata} 
%\end{table} 
The observation process described above with the example provides us with two problems that need to be addressed simultaneously: 
(1) Determining whether an animal passes through the detector (trap station) in the face of imperfect detection of detectors. 
(2) Reconciling partially-identified individuals.
While the second problem has been recently tackled in \cite{royle2015spatial}, our emphasis in this paper is to tackle the first and integrate it into the solution of the second. 
%\paragraph{Synchronisation of bilateral detection histories.}\label{synchronisation}
% \subsubsection{Problem definition}
\subsection{Model development}\label{model_development}
\subsubsection{Disentangling animal entrance and detection in spatial capture-recapture models}\label{gpid}
We note that for an animal to be observed by a detector at a given location and occasion, the animal has to (1) pass through the trap station housing the detector and (2) has to be captured by the detector given that the animal has passed through. We aim to disentangle these two processes by utilizing, once again, the hierarchical modelling approach. From Example \ref{example:sampledata}, there are four types of detection histories observable at a given trap station on a given sampling occasion:  `11' (observed by both detectors), `10' (observed by detector 1 but not by detector 2), `01' (not observed by detector 1 but observed by detector 2) and `00' (not observed by either detector). The first three histories (`11', `10' and `01') conclusively state that the animal passed through the trap station since we have one observation. But in the fourth case (`00'), we are presented with two possibilities: (a) the animal passed through the trap station and both detectors failed to record this event or (b) the animal did not pass through the trap station. 
% the probability that an individual has detection history `$10$' by two independent collocated detectors on some particular occasion will be $\pi p_1(1-p_2)$, where $\pi$ denotes the probability of animal entrance in the vicinity of the detectors and $p_1$, $p_2$ denotes the probability of detection by detector 1 and 2 conditional on animal entry, respectively. Note that, nondetection of an individual in any of the detectors does not necessarily imply absence - it rather means, either the species was present and was not detected in any of the detectors, or the individual was not present. The probability of an individual having detection history `$00$' at the same trapping location on some particular occasion will be $\pi (1-p_1)(1-p_2) + (1-\pi)$. We shall go into more details of modelling in the next section. 
% \subsection{Spatial capture-recapture model for disentangling animal entrance and detection}
\paragraph{Defining the state process of animal entry to trap station}
Let $E_{ikt}$ be a latent variable that indicates whether individual $i$ has entered a trap station $\mathbf{u_k}$ on occasion $t$ $\left(E_{ikt}=1\right)$ or not $\left(E_{ikt} = 0\right)$. Further, let $\pi_{ik} = P(E_{ikt} = 1)$ be the probability of the corresponding event of trap entrance. We model the probability that an individual $i$ passes through a trap station $\mathbf{u_k}$ as a decreasing function of distance between its activity centre $\mathbf{s_i}$ and trap station $\mathbf{u_k}$. A typical model to describe `trap entry probability' $\pi_{ik}$ is the Gaussian form of the type $\pi_{ik} = p_0\, \exp(-d_{ik}^2/(2\,\sigma^2) )$, 
where $d_{ik} = d(\mathbf{s_i}, \mathbf{u_k}) = \Arrowvert \mathbf{s_i} - \mathbf{u_k} \Arrowvert$ is the Euclidean distance between points $\mathbf{s_i}$ and $\mathbf{u_k}$, $p_0$ is called `baseline trap entry probability' and $\sigma$ quantifies the rate of decline in trap entry probability as the distance between individual activity centre $\mathbf{s_i}$ and trap station $\mathbf{u_k}$ increases. We note with interest, that previous SECR models regard this modelling structure as part of the observation process such that $p_0$ is instead regarded as the `baseline encounter probability' and $\sigma$ is instead regarded as the rate of decline in detection probability as the distance between individual activity centre $\mathbf{s_i}$ and trap station $\mathbf{u_k}$ increases (\citealt{borchers2008spatially}, \citealt{royle2009bayesian}).

We proceed with the Gaussian form in our development, while recognizing that there can be many other options to define the rate of decline in animal trap entry probability to represent other realities. Further, it is often the case that gender acts as an important covariate to define the extent of animal movement \citep{sollmann2011improving}. For example, often males and females have different extents of spatial movement, defined by the parameter $\sigma$ in our development. We then define $\sigma$ as the following: 
%In practice some of the elements of $\mathbf{x}$ may be unobserved. Let $\mathbf{x_{obs}}$ be the part of $\mathbf{x}$ which is observed and $ \mathbf{x_0}$ be the other part which is missing. 
$\sigma (x_i) = \sigma_m$,  if $x_i=1$, i.e., individual $i$ is a male; $\sigma (x_i) = \sigma_f$,  if $x_i=0$, i.e., individual $i$ is a female. Here each $x_i$ is assumed to follow the Bernoulli distribution with parameter $\theta$, $\theta$ being the probability that an arbitrary individual in the population is male. Additionally, the explicit recognition of these sex effects will, later, be very helpful in synchronising the partially identified individuals as seen in Example \ref{example:sampledata} because we can utilize the fact that sex is ascertained for each individual $i$ and we constrain the sychronisation to probabilistically linking partially identified individuals of only the same sex. 
%We assume each $x_i$ independently follows Bernoulli distribution with parameter $\theta$, $\theta$ being the probability that an arbitrary individual in the population is male. Thus, this structure will also permit us to estimate sex ratio - an ecological quantity of much interest. %random vector where $\mathbf{x_{obs}}$ and $\mathbf{x_0}$ are the set of observed and unobserved components of $\mathbf{x}$, respectively. 
%the $i^{\text th}$ component of  $\mathbf{x} = (\mathbf{x_{obs}}, \mathbf{x_0})$, 
\paragraph{Defining the observation process at trap stations}
Here, we introduce the detection probabilities for our observation model conditional on the entry at a trap station. Let $p^{(12)}_k$ be the probability of detection by both detectors simultaneously at $\mathbf{u_k}$ on a sampling occasion, $p^{(1\overline{2})}_k$ be the probability of detection only by detector 1 at $\mathbf{u_k}$ on a sampling occasion and $p^{(\overline{1}2)}_k$ be the probability of detection only by detector 2. Collection of these different probabilities is denoted by $\mathbf{p^{(12)}} = (p^{(12)}_1, p^{(12)}_2, \dots, p^{(12)}_K)$, 
$\mathbf{p^{(1\overline{2})}} = (p^{(1\overline{2})}_1, p^{(1\overline{2})}_2, \dots, p^{(1\overline{2})}_K)$,
$\mathbf{p^{(\overline{1}2)}} = (p^{(\overline{1}2)}_1, p^{(\overline{1}2)}_2, \dots, p^{(\overline{1}2)}_K)$ and 
$\mathbf{p} = (\mathbf{p^{(12)}},\, \mathbf{p^{(1\overline{2})}},\, \mathbf{p^{(\overline{1}2)}})$.
As described in Section~\ref{crdata}, $y_{ikt}^{(1)}$, $y_{ikt}^{(2)}$ are binary responses corresponding to detections on detectors 1 and 2 respectively, defined only when $E_{ikt} =1$. However, when $E_{ikt} = 0$, both $y_{ikt}^{(1)}$, $y_{ikt}^{(2)}$ have degenerate distributions at 0.
%, i.e., $f(y_{ikt}^{(1)}, y_{ikt}^{(2)} \, | \, \mathbf{p}, E_{ikt} =0) = I [(y_{ikt}^{(1)}, y_{ikt}^{(2)}) = (0,0)]$, for an individual $i$ failing to pass through the trap station $\mathbf{u_k}$ and is not susceptible for observation by either detector 1 or 2. Here, $I(\emph{condition})$ is an indicator function, which takes the value 1 if $\emph{condition}=\mathrm{TRUE}$ and takes the value 0 otherwise. In this case, $I [(y_{ikt}^{(1)}, y_{ikt}^{(2)}) = (0,0)]$ will necessarily take the value 1 if $E_{ikt} = 0$. 
\subsubsection{Development of the joint posterior density of all parameters} \label{joint_posterior}
\paragraph{Joint posterior density to disentangle animal trap entry and trap detections} When $E_{ikt} = 1$, the conditional probability of detection in detector 1 is $p^{(1)}_k=p^{(12)}_k+p^{(1\overline{2})}_k$ and the conditional probability of detection in detector 2 is $p^{(2)}_k=p^{(12)}_k+p^{(\overline{1}2)}_k$. If we assume both the detectors are of same quality, $p^{(1)}_k=p^{(2)}_k =\phi_k$ for each $k$. Consequently, we have $p^{(1\overline{2})}_k=p^{(\overline{1}2)}_k=\phi_k - p^{(12)}_k$. We note here that, $E_{ikt} = 1$ if $ y_{ikt}^{(1)} + y_{ikt}^{(2)} >0$. In contrast, $E_{ikt}$ is unobserved if $(y_{ikt}^{(1)}, y_{ikt}^{(2)}) = (0,0)$.  
%For example, suppose that individual $i$ gets detected in detector 1 at location $\mathbf{u_k}$ on occasion $t$. Then individual $i$ may either get detected in detector 2 or may not. 
%Note that, detecting left or right side of the individual depends on the alignment of individual, not the camera devices. %Further formulation of the observation model is developed under this assumption. 
%We use the following notation for the detection probabilities for the rest of the article : $p^{(1)}=p^{(2)}=\phi$. As a simple example for illustrating the bilateral detection probabilities let us assume that $k=1$ and $t=1$. 
We can then construct probability arguments to compute probabilities of various data outcomes for an individual $i$ at trap station $\mathbf{u_k}$ on sampling occasion $t$ : 
(i) $P[(y_{ikt}^{(1)}, y_{ikt}^{(2)}) = (1,0)] =$ $\pi_{ik} (\phi_k - p^{(12)}_k) =$ $P[(y_{ikt}^{(1)}, y_{ikt}^{(2)}) = (0,1)]$, 
(ii) $P[(y_{ikt}^{(1)}, y_{ikt}^{(2)}) = (1,1)] =$ $\pi_{ik}  p^{(12)}_k$, 
(iii) $P[(y_{ikt}^{(1)}, y_{ikt}^{(2)}) = (0,0)] =$ $(1-\pi_{ik}) + \pi_{ik} (1-(2 \phi_k - p^{(12)}_k))$.
%\begin{align*}
%&\text{(i)} \hspace{5pt} P[(y_{ikt}^{(1)}, y_{ikt}^{(2)}) = (1,0)] = P[E_{ikt} = 1] \, P[(y_{ikt}^{(1)}, y_{ikt}^{(2)}) = (1,0) \, | \, E_{ikt} = 1]  \\
%&\hspace{118pt} = \pi_{ik} (\phi_k - p^{(12)}_k) = P[(y_{ikt}^{(1)}, y_{ikt}^{(2)}) = (0,1)],\\
%&\text{(ii)} \hspace{5pt} P[(y_{ikt}^{(1)}, y_{ikt}^{(2)}) = (1,1)] = P[E_{ikt} = 1] \, P[(y_{ikt}^{(1)}, y_{ikt}^{(2)}) = (1,1) \, | \, E_{ikt} = 1] = \pi_{ik}  p^{(12)}_k,\\
%&\text{(iii)} \hspace{5pt} P[(y_{ikt}^{(1)}, y_{ikt}^{(2)}) = (0,0)]= P[E_{ikt} = 0] \, P[(y_{ikt}^{(1)}, y_{ikt}^{(2)}) = (0,0) \, | \, E_{ikt} = 0] \,  + \\
%&\hspace{170pt}  P[E_{ikt} = 1] \, P[(y_{ikt}^{(1)}, y_{ikt}^{(2)}) = (0,0) \, | \, E_{ikt} = 1] \\
%&\hspace{125pt} = (1-\pi_{ik}) + \pi_{ik} (1-(2 \phi_k - p^{(12)}_k)) = 1 - \pi_{ik}(2 \phi_k - p^{(12)}_k).
%\end{align*}
Note that, $E_{ikt} = 1$ in the first two cases (i) and (ii), as $y_{ikt}^{(1)} + y_{ikt}^{(2)} >0$ in both the cases. But in the third case (iii), we are unsure of $E_{ikt}$ because it is unobserved as the individual $i$ has not been observed by either detector 1 or 2. If we assume that both the detectors function independently of each other, then $p^{(12)}_k = p^{(1)}_k p^{(2)}_k = \phi_k^2$ and $p^{(1\overline{2})}_k = p^{(\overline{1}2)}_k = \phi_k(1-\phi_k)$, for each $k$. The data outcomes above then become :
(i)  $P[(y_{ikt}^{(1)}, y_{ikt}^{(2)}) = (1,0)] =$   $\phi_k(1-\phi_k) \pi_{ik} =$ $P[(y_{ikt}^{(1)}, y_{ikt}^{(2)}) = (0,1)]$,
(ii) $P[(y_{ikt}^{(1)}, y_{ikt}^{(2)}) = (1,1)] =$ $\phi_k^2 \, \pi_{ik}$, 
(iii) $P[(y_{ikt}^{(1)}, y_{ikt}^{(2)}) = (0,0)] =$ $(1-\pi_{ik}) + (1-\phi_k)^2 \pi_{ik}= 1 -\phi_k(2-\phi_k) \pi_{ik}$.
Note that the population size $N$, which is a parameter of major interest, is an unknown quantity. Due to this, the number of some other variables including some latent variables is unknown and therefore the dimension of the parameter space is also unknown. This is one of the main difficulties in analysing the proposed SECR model. We consider the method of data augmentation \citep{royle2009bayesian} for analysing the proposed SECR model to handle this difficulty. This is implemented by choosing a large integer $M$ to bound $N$ and augmenting the two observed data sets with a large number of ``all-zero'' encounter histories.
%(corresponding to those individuals which do not exist in the population).
 We denote the zero-augmented data sets 
% (with all-zero detection history)
  by $\mathbf{Y^{(1)}}$ and $\mathbf{Y^{(2)}}$, corresponding to detectors 1 and 2 respectively; each of these is now of dimension $M\times K \times J$. A vector of $M$ latent binary variables $\mathbf{z}$ is introduced to account for the zero-inflation in the data sets and each $z_i$. 
In other words, these variables indicate which individuals are present in the population and are 
%is 
modelled with the Bernoulli distribution with parameter $\psi$. Thus, the true population size 
%$N \sim \mathrm{Binomial}\left(M,\psi\right)$
$N$ follows the Binomial distribution with parameters $M$ and $\psi$ 
(see Section~\ref{stateprocess}).
% (note this definition in the description of the State Process). 
%Consequently, the data augmented individuals who are present in the population (i.e with $z_i = 1$), are also allocated independent activity centres over the state space $\mathcal{V}$.
%\MakeUppercase{We have also augmented the array of activity centres and keep the same notation  $\mathbf{S}$ for convenience (dimension of augmented $\mathbf{S}$ is $M \times 2$).}
%We consider the method of parameter-expanded data augmentation (\citealt{royle2012parameter}) for analysing the SECR model developed here. We denote the zero-augmented data sets (with all-zero detection history) by the vectors $\mathbf{Y^{(1)}}$ and $\mathbf{Y^{(2)}}$, corresponding to detectors 1 and 2 respectively; each of which is of dimension $M\times K \times J$. A vector of $M$ latent binary variables $\mathbf{z}$, is introduced to indicate which individuals are truly in the population with $z_i \sim \text{Bernoulli} (\psi)$, inducing the relationship with abundance size $N \sim \text{Binomial} (M, \psi)$ (note this definition in the description of the State Process). 
%We also augment the covariate sample on genders $\mathbf{x_{obs}}$ with the unobserved Bernoulli($\theta$) variables $\mathbf{x_0}$
%$ = (x_{n+1}, x_{n+2}, \dots, x_M)$
The augmented latent vector on sex category is denoted by $\mathbf{x}$. Let $\mathbf{x_{obs}}$ be a vector of binary observation (length $n \times 1$) on sex category of the captured individuals
% and the vector of latent missing observations in $\mathbf{x}$ be denoted by $\mathbf{x_0}$ (length $(M-n) \times 1$).
: $x_i \, (\in \mathbf{x_{obs}})$ takes the value 1 when individual $i$ is a male, takes the value 0 if its female. The vector of latent missing observations in $\mathbf{x}$ is denoted by $\mathbf{x_0}$ (length $(M-n) \times 1$). 
% which is assumed to be partially observed. 
Assuming that detection histories coming from detector 1 and 2 are in synchronised order and covariate information (partially observed) on individual sex category is available for each real individual (with $z_i = 1$), the joint density of $( \mathbf{Y^{(1)}}, \mathbf{Y^{(2)}} ) = ((y_{ikt}^{(1)}, y_{ikt}^{(2)}))_{i,k,t}$
%$ \mathbf{Y} = ( \mathbf{Y^{(1)}}, \mathbf{Y^{(2)}} ) = ((y_{ikt}^{(1)}, y_{ikt}^{(2)}))_{i,k,t}$
 and $\mathbf{x}$ is the following: 
\begin{align}
& f(\mathbf{Y^{(1)}}, \mathbf{Y^{(2)}}, \mathbf{x_{obs}} \, | \, \mathbf{z}, \mathbf{x_0}, \theta, {\boldsymbol \phi}, p_0, \sigma_m, \sigma_f, \mathbf{S})
% = \prod_{i=1}^M \Big{[} f (\mathbf{Y^{(1)}_{i, obs}}, \mathbf{Y^{(1)}_{i, obs}} \, | \, z_i, {\boldsymbol \phi}, p_0, \sigma_m, \sigma_f, \mathbf{s_i}) \, g(\mathbf{x_i}\, | \, z_i, \theta ) \Big{]} \nonumber\\
= \prod_{i=1}^{M} \Big{[} \prod_{k=1}^{K} \, \prod_{t=1}^{J} \, \big{\{} (1-\pi_{ik}) \, I(y_{ikt}^{(1)} + y_{ikt}^{(2)} =0) + \nonumber\\
&  \pi_{ik} \, \phi_k^{(y_{ikt}^{(1)} + y_{ikt}^{(2)})} \, (1-\phi_k)^{2-(y_{ikt}^{(1)} + y_{ikt}^{(2)})} \big{\}}^{z_i} \, \theta^{z_i x_i} (1-\theta)^{z_i(1-x_i)} \Big{]},  
%&=\prod_{i=1}^{M} \,
%\prod_{k=1}^{K} \prod_{t=1}^{J} \,  \Big{\{}\pi_{ik} \phi_k^{(y_{ikt}^{(1)} + y_{ikt}^{(2)})} (1-\phi_k)^{2-(y_{ikt}^{(1)} + y_{ikt}^{(2)})} \Big{\}}^{z_i I( y_{ikt}^{(1)} + y_{ikt}^{(2)} >0)} 
%\Big{\{}(1-\pi_{ik}) +  \pi_{ik} \, (1-\phi_k)^2 \Big{\}}^{ z_i I(y_{ikt}^{(1)} + y_{ikt}^{(2)} =0)} \nonumber \\
%& \hspace{20pt} \times \prod_{i=1}^{M} \theta^{z_i x_i} (1-\theta)^{z_i(1-x_i)},
\label{intlik.lr}
\end{align} 

It is straightforward to handle the latent missing observations in $\mathbf{x}$, denoted by $\mathbf{x_0}$, using a Bayesian MCMC analysis \citep{royledorazio2008hierarchical}.
% \citep{royle2015likelihood}. 
%(CITE ROYLE ET AL - LIKELIHOOD FORMULATION OF CATEGORICAL DATA). 
For simplicity, we can assume $\phi_k = \phi$, for each $k$.
The posterior density of parameters $\{\mathbf{z}$, $\mathbf{x_0}$, $\psi$, $\theta$, $\phi$, $p_0$,  $\sigma_m$, $\sigma_f$, $\mathbf{S} \}$ can be obtained as follows:%
\begin{align}
& g (\mathbf{z}, \mathbf{x_0}, \psi, \theta, \phi, p_0,  \sigma_m, \sigma_f, \mathbf{S} \mid \, \mathbf{Y^{(1)}}, \mathbf{Y^{(2)}}, \mathbf{x_{obs}}) \nonumber\\
& \propto f(\mathbf{Y^{(1)}}, \mathbf{Y^{(2)}}, \mathbf{x_{obs}} \, | \, \mathbf{z}, \phi, p_0,  \sigma_m, \sigma_f,\mathbf{S}) \, g(\mathbf{x_0}\, |\,\mathbf{z}, \theta) \, g(\mathbf{z}\, |\, \psi) \, g( \psi, \theta, \phi, p_0, \sigma_m, \sigma_f, \mathbf{S}) \nonumber\\
&  = \prod_{i=1}^{M} \Big{[} \prod_{k=1}^{K} \, \prod_{t=1}^{J} \, \big{\{} (1-\pi_{ik}) \, I(y_{ikt}^{(1)} + y_{ikt}^{(2)} =0) +  \pi_{ik} \, \phi^{(y_{ikt}^{(1)} + y_{ikt}^{(2)})} \, (1-\phi)^{2-(y_{ikt}^{(1)} + y_{ikt}^{(2)})} \big{\}}^{z_i}   \nonumber \\
& \hspace{50pt} \times \, \theta^{z_i x_i} (1-\theta)^{z_i(1-x_i)} \, \psi^{z_i} (1-\psi)^{1-z_i} \Big{]}\, \times \, g(\psi, \theta, \phi, p_0, \sigma_m, \sigma_f, \mathbf{S}),
%& \hspace{10pt} =\prod_{i=1}^{M}\,\Big{[} \, \Big{\{}
%\psi \, \theta^{x_i} (1-\theta)^{(1-x_i)} \, \phi^{(y_{i\cdot \cdot}^{(1)} + y_{i\cdot \cdot}^{(2)})} (1-\phi)^{2n_{i\cdot}-(y_{i\cdot \cdot}^{(1)} + y_{i\cdot \cdot}^{(2)})} \,
%\prod_{k=1}^{K} \,  \pi_{ik}^{n_{ik}} \{(1-\pi_{ik}) + \pi_{ik} (1-\phi)^2 \}^{J-n_{ik}}\Big{\}}^{z_i} \nonumber\\
%& \hspace{50pt} \times  (1-\psi)^{1-z_i} \Big{]}\, \times \, g(\psi, \phi, p_0, \sigma_m, \sigma_f, \mathbf{S}),
\label{posterior.lr}
\end{align} where $g(\psi, \theta, \phi, p_0, \sigma_m, \sigma_f, \mathbf{S})$ is the prior density for the parameters $\psi, \theta, \phi, p_0, \sigma_m, \sigma_f,\mathbf{S}$; $g(x_0|z,\theta)$ is the conditional prior density, which is that of the Bernoulli distribution with parameter $\theta$ when $z$ takes the value 1, and $g(z|\psi)$ is also the conditional prior density, which is that of the Bernoulli distribution with parameter $\psi$.
%(ALSO MENTION WHAT $g(x_0|z,\theta)$ AND $g(z|\psi)$ ARE)\\
%(\textsc{Thanks. I have mentioned them above.})
\paragraph{Joint posterior density to include bilateral synchronisation complexity} \label{synchronisation}
The relationship (\ref{posterior.lr}), however, does not deal with the problem of synchronising data from the partially identified individuals as described in Table~\ref{sampledata}. Ignoring this problem will result in overestimation of abundance, underestimation of standard errors and poor coverage for credible interval estimates \citep{bonner2013mark}. 
We integrate the solution used by \cite{royle2015spatial} into our problem formulation (\ref{posterior.lr}). 

Accordingly, the two lists of capture histories generated as in Table~\ref{sampledata} essentially come from the same population and therefore there must be a unique association between the two lists. As noted earlier, we are particularly interested to form the associations for the `partially identified' individuals. Accordingly, we treat the true identity of a partially identified individual as a latent variable. We then probabilistically link individuals from the two lists obtained from detector 1 and detector 2, respectively, by introducing a latent identity variable $\mathbf{L}=(\text{L}_1, \text{L}_2, \dots, \text{L}_M)^\prime$. $\mathbf{L}$ is a permutation of $\{1,2,\dots,M\}$ which re-orders the set of individuals from detector 2 to correspond with the set of individuals from detector 1.

%$\mathbf{L}$ is a one to one mapping from an index set of individuals from detector 2 to $\{1,2,\dots,M\}$ providing the true index of each detector 2 individuals.\\
%(I DON'T UNDERSTAND THIS SENTENCE.)\\
%(\textsc{Thanks. I have rewritten the sentence.})

More details on the synchronisation procedure can be found in \cite{royle2015spatial}. Without loss of generality, we define the true identity of each individual in the population to be in the row-order of capture histories of detector 1. Then we reorder the rows of detector 2 data set $\mathbf{Y^{(2)}}$ as indicated by $\mathbf{L}$ to synchronise with the individuals of the detector 1 data set $\mathbf{Y^{(1)}}$. We denote this newly ordered detector 2 data set as $\mathbf{Y^{(2*)}}$. Now these two synchronized data sets can be used in the SECR model (\ref{posterior.lr}).
%Conditional on latent variable $\mathbf{L}$, the proposed model will reduce to the spatial capture-recapture model defined in the relationship (\ref{posterior.lr}). 
%(I DON'T UNDERSTAND THIS SENTENCE.)\\
%(\textsc{Thanks. I have rewritten the sentence.})
 An individual $i$ will be called `detected' if there exists a non-zero observation $y_{ikt}^{(1)}$ or $y_{ikt}^{(2*)}$ for some $k$ and $t$; that is, if $y_{i\cdot \cdot}^{(1)}+y_{i\cdot \cdot}^{(2*)}>0$. Thus, if we obtain detection history observations $\mathbf{Y_{obs}^{(1)}}$ and $\mathbf{Y_{obs}^{(2)}}$ from two detectors during a spatial capture-recapture survey, they may not be synchronised (see Example~\ref{example:sampledata}). 
Our aim will be to use the latent vector $\mathbf{L}$ to synchronise $\mathbf{Y_{obs}^{(1)}}$ and $\mathbf{Y_{obs}^{(2)}}$. 
%Covariate information on individual sex category is helpful in synchronising such bilateral data, because the gender of every individual is unique. 
Accordingly, by integrating this synchronisation complexity into the joint posterior density (\ref{posterior.lr}), we obtain the new combined posterior of parameters $\{\mathbf{z}, \mathbf{x_0}, \psi, \theta, \phi, p_0, \sigma_m, \sigma_f,\mathbf{S}, \mathbf{L} \}$ as follows:
\begin{align} 
& g ( \mathbf{z}, \mathbf{x_0}, \psi, \theta, \phi, p_0, \sigma_m, \sigma_f, \mathbf{S}, \mathbf{L} | \mathbf{Y^{(1)}}, \mathbf{Y^{(2*)}}, \mathbf{x_{obs}})\propto f(\mathbf{Y^{(1)}}, \mathbf{Y^{(2*)}}, \mathbf{x_{obs}} \, | \, \mathbf{z}, \mathbf{x_0}, \psi, \theta, \phi, p_0, \sigma_m, \sigma_f,\mathbf{S}, \mathbf{L})  \nonumber \\ 
&  \times \,  g (\mathbf{z} | \psi) \, g (\mathbf{x_0}\, | \, \theta) \, g ( \psi, \theta, \phi, p_0, \sigma_m, \sigma_f,  \mathbf{S}, \mathbf{L}) = \prod_{i=1}^{M} \Big{[} \prod_{k=1}^{K} \, \prod_{t=1}^{J} \, \big{\{} (1-\pi_{ik}) \, I(y_{ikt}^{(1)} + y_{ikt}^{(2*)} =0) \nonumber\\
&  +  \pi_{ik} \, \phi^{(y_{ikt}^{(1)} + y_{ikt}^{(2*)})} \, (1-\phi)^{2-(y_{ikt}^{(1)} + y_{ikt}^{(2*)})} \big{\}}^{z_i}   \, \theta^{z_i x_i} (1-\theta)^{z_i(1-x_i)} \, \psi^{z_i} (1-\psi)^{1-z_i} \Big{]}\, \times \, g(\psi, \theta, \phi, p_0, \sigma_m, \sigma_f, \mathbf{S}) \nonumber\\
&  =\prod_{i=1}^{M}\,  \Big{[} \Big{\{} \psi \theta^{x_i} (1-\theta)^{1-x_i} \,
\phi^{(y_{i\cdot \cdot}^{(1)} + y_{i\cdot \cdot}^{(2*)})} (1-\phi)^{2n_{i\cdot}-(y_{i\cdot \cdot}^{(1)} + y_{i\cdot \cdot}^{(2*)})} \,
\prod_{k=1}^{K} \,  \pi_{ik}^{n_{ik}} \{(1-\pi_{ik}) + \pi_{ik} (1-\phi)^2 \}^{J-n_{ik}}\Big{\}}^{z_i} \nonumber\\
& \times  (1-\psi)^{1-z_i} \Big{]}\, \times \, g (\psi, \theta, \phi, p_0, \sigma_m, \sigma_f, \mathbf{S}, \mathbf{L}),
\label{posterior.lr.A}
\end{align}where $n_{ik}=\sum_{t=1}^J I(y_{ikt}^{(1)} + y_{ikt}^{(2*)} >0)$ and $g (\psi, \theta, \phi, p_0, \sigma_m, \sigma_f, s, \mathbf{L})$ is the joint prior density for the parameters $\psi, \theta, \phi, p_0, \sigma_m, \sigma_f, \mathbf{S}, \mathbf{L}$.
The MCMC algorithm used to sample from this posterior density (\ref{posterior.lr.A}) is detailed in Web Appendix B. %
%=============================================================================
\subsubsection{Identifiability of model parameters} \label{identifiability}
It is necessary to check for issues of identifiability when new models and estimators such as ours are proposed. Inherent identifiability issues in the model give rise to problems of variance inflation, estimation biases and also false specification of the number of true parameters in penalized methods of model selection \citep{gimenez2003parameter}.
%If there are inherent identifiability issues during the construction of the model itself, we may have problems of variance inflation, estimation biases and also false specification of the number of true parameters in penalized methods of model selection \citep{gimenez2003parameter}. 
We evaluate the identifiability concerns of two important pairs of parameters in our SECR model, $\left(\phi,\pi\right)$ and $\left(p_0,\sigma\right)$. 
%\section{Identifiability of different parameters} \label{identifiability}
\paragraph{Identifiability between $\phi$ and $\pi$}  \label{identifiability:phi_pi}
The relevant probability statements describing the probability of the data conditional on the parameters for detection probability $\phi$ of a detector and the trap entry probability $\pi$, is given by 
\begin{align}
	f&(y^{(1)}, y^{(2)} \, | \, \phi, \pi)= (1-\pi)\, I(y^{(1)} + y^{(2)}= 0) +\pi\,
	\phi^{(y^{(1)} + y^{(2)})} (1-\phi)^{2-(y^{(1)} + y^{(2)})}.
	\label{eq:intlik.identifiability1}
	\end{align}
Note that equation (\ref{eq:intlik.identifiability1}) is a four cell multinomial model, where the cells are `00', `01', `10' and `11'. This model is identifiable, provided both $\phi$ and $\pi$ lie strictly between 0 and 1. The formal proof is derived in Web Appendix A.1.
%~\ref{app:identifiability:phi_pi}. 
However, even with this condition, it is always possible that the given data (mostly due to inadequate sample size) may appear to only arrive in the form of `11' and `00' pairs. In such a case as well, we will have issues of non-identifiability. 
%The presence of partial capture (0,1) or (1,0) is key component in the data set.
%\noindent ---------------------------------------------------------------------------------------------------------------------
%=================================================================
\paragraph{Identifiability between $p_0$ and $\sigma$} \label{identifiability:p0_sigma}
The trap entry probability $\pi$ is modelled as a decreasing function of distance between location of activity centre of an individual and a trap station. The two parameters in the model for trap entry probability $\pi$ are: (a) the baseline trap entry probability parameter $p_0$ and (b) the scale parameter $\sigma$. 
This pair of parameters  $\left(p_0,\sigma\right)$ is identifiable under the condition that there exist two observation indices $(i_1, k_1)$ and $(i_2, k_2)$ such that $z_{i_1}>0$, $z_{i_2}>0$ and $d(\mathbf{s_{i_1}}, \mathbf{u_{k_1}}) \neq d(\mathbf{s_{i_2}}, \mathbf{u_{k_2}})$. Here, $(i,k)$ represents the indices of the pair individual ($i$), trap station ($k$). 
It is sufficient if the index of $(\mathbf{s_{i_1}}, \mathbf{u_{k_1}})$ is different from the index of $(\mathbf{s_{i_2}}, \mathbf{u_{k_2}})$, implying that we achieve identifiability if $\mathbf{s_{i_1}} \neq \mathbf{s_{i_2}}$ or $\mathbf{u_{k_1}} \neq \mathbf{u_{k_2}}$ or both, $\mathbf{s_{i_1}} \neq \mathbf{s_{i_2}}$ and $\mathbf{u_{k_1}} \neq \mathbf{u_{k_2}}$ as long as $d(\mathbf{s_{i_1}}, \mathbf{u_{k_1}}) \neq d(\mathbf{s_{i_2}}, \mathbf{u_{k_2}})$. This condition is proved in Web Appendix A.2.
%~\ref{app:identifiability:p0_sigma}. 
\subsubsection{Posterior Propriety}
\cite{link2013cautionary} brought up an important and often overlooked aspect of posterior impropriety during Bayesian analysis of estimation problems in ecology and stresses the need for practitioners to ensure that posteriors are proper.
%With respect to problems in ecology, \cite{link2013cautionary} brought up an important and often overlooked aspect of posterior impropriety during Bayesian analyses of estimation problems and made a persuasive case that practitioners ensure that posteriors be proper.
More recently, \cite{gopalaswamy2016examining} indirectly suggest the use of defensibly informed or bounded priors to ensure posterior propriety in such problems and indicate the close association between posterior impropriety and identifiability. Accordingly, in this paper, we implement bounded priors based on ecologically justifiable upper limits for all the parameters used in our model. The assumed proper prior distributions for these parameters along with other model parameters and latent variables are as follows: a uniform distribution over the interval $(0,1)$ for the probability parameters $\phi$, $p_0$, $\psi$ and $\theta$; a uniform distribution over the interval $(0,R)$ for parameters $\sigma_m$ and $\sigma_f$ where $R$ is high enough to expect that it would be impossible for animals to exhibit movement as widely as this scale during sampling. $\mathbf{L}$ has a Uniform distribution over the permutation space of $\{1,2,\dots,M\}$. Each $z_i$ follows a Bernoulli$(\psi)$ distribution and each $x_i$ follows a Bernoulli$(\theta)$ distribution. Each $\mathbf{s_i} = (s_{i1}, s_{i2})^\prime$ follows uniform distribution over the state space $(\mathcal{V})$. All the parameters are distributed independently of each other. %
\subsubsection{Use of covariates}\label{covariates} 
The advantage of the estimator we have developed in this study will only be realized effectively if trap-specific covariates are provided as explanatory variables for the ecological process parameter, $p_0$, as well as observation process parameter, $\phi$. In practical wildlife surveys using camera traps \citep{Oconnell2011camera}, 
investigators may be interested to assess the movement ecology of animals and assess what factors drive animals to visit particular trap stations or not. For example, investigators might be interested to test the effectiveness of various lures/baits at trap stations or identify local site characteristics that attract or repel animals. These explanatory variables may suitably describe the variation in trap entry probability, $p_0$. However, such covariates are likely to have little influence on whether the cameras installed at trap stations work effectively or not. Instead some other covariates may better describe factors influencing how well the cameras fire and capture records of animals passing by. For example, the camera trap brand and time of the day (especially in passive infra-red cameras) may influence how well the cameras fire. Hence, such covariates can adequately describe the detection probability of the detector, $\phi$. It is common practice in ecology, to permit for such covariates ($\mathbf{h}$) in the model using a logit-link
% (see eq.~\ref{fncov} below)
for $p_0$ and/or $\phi$. 
\subsection{Assessment of model performance}
\subsubsection{Simulation Design}\label{simstudy}
Admittedly, for a high dimensional problem such as this, it would be infeasible to assess model performance for an exhaustive range of parameters simply owing to the number of combinations and computation time. We conducted simulations for 70 scenarios (provided in Web Table~1) grouped into 2 equal sized sets, to assess the performance of the model proposed here. We set $\sigma_m$ = 0.3 and $\sigma_f$ = 0.15 for the first set of 35 scenarios, $\sigma_m$ = 0.4 and $\sigma_f$ = 0.2 for the second set of 35 scenarios. The simulation design was aimed to highlight the importance of identifying the pair of parameters ($p_0$ and $\phi$) and its effect on the robustness of estimates of other parameters (especially $N$). We set $p_0$  = 0.005, 0.01, 0.03, 0.05, 0.07; $\phi$ = 0.3, 0.4, 0.5, 0.6, 0.7, 0.8, 0.9, which gives us 35 different scenarios for each of the two sets corresponding to the values taken by  $p_0$ and $\phi$. We assumed that a total of 100 individuals are residing inside the state space of which 40 are male. Each of the simulation experiments is conducted within a rectangular state space of dimension 5 unit $\times$ 7 unit (Web Figure~1), after setting a buffer of 1 unit in both horizontal and vertical directions, a $10 \times 16$ trapping array of total 160 trap stations has been set (trap spacing is 0.3 unit on $X$ axis and 0.3125 unit on $Y$ axis). Each of the traps remains active for $J=50$ sampling occasions simultaneously. 
For parameter estimation, we set the maximum possible number of individuals present in the population ($M$) at 400 
%(this includes the data augmentation number) 
for all the scenarios. The MCMC chains for each of the parameters are obtained (each of length 30000) and the estimates were computed using those chains with a burn-in of 10000. 
%The aim of the simulation study is to demonstrate (1) that an increase in information content (of increased knowledge of full detection histories known) will lead to better estimates and (2) that incorporating gender information will lead to better estimates.
%-----------------------------------------------------------
%Let us first look at the scatter plots between $p_0$ and $\phi$ for each of the 70 scenarios (Figures~\ref{scatplot135}, \ref{scatplot3670}). 
%As identifiability between $p_0$ and $\phi$ is very much similar with the identifiability between occupancy probability and detection probability in case of occupancy models \citep{mackenzie2002estimating}. For occupancy probability and detection probability to be identifiable the number of surveyed sites and/or the number of detections in each surveys are required to be more than 1. 
%$p_0$has wider spread with 95\% CI width $<0.03$ for small values of $\phi = (0.3, 0.4)$ and posterior distributions of $\phi$ has wider spread with 95\% CI width $<0.24$ for small values of $p_0 = (0.005, 0.01)$. 
%on the identifiability issue between $p_0$ and $\phi$ using these two scenarios will not be correct. Because of these, data sets with very low number of recaptures resulted in biased estimates with large RMSEs and very wide 95\% CIs. %Number of trap entries and/or number of recaptures in each of the individuals are required to be more than 1 for the two parameters $p_0$ and $\phi$ to be identifiable. 
%( except scenario 45, where posterior correlation was -0.2 which we believe to be a sampling fluctuation).
\subsubsection{Comparison with `unidentified' model}\label{RoylePartialID}
Often practitioners are interested to know about robustness of estimates of particular parameters of interest under violations of model assumptions. For example, ecologists are very interested in $N$ and will often base the choice of their models based on robustness of estimates of $N$ in the face of model violations. Motivated by this concern we also performed a parallel simulation study by collapsing the two parameters ($\phi$ and $p_0$) into one parameter. In effect, the model reduces to the partial identification model proposed by \cite{royle2015spatial}.
%(CITE ROYLE2015PAPER ON PARTIAL ID) We have also performed a simulation study for comparing estimates with Royle's partial ID model.
Table~\ref{compare2} provides an illustrative example to demonstrate the need for practitioners to use the model we have proposed in this paper by indicating the biases in estimates of $N$ and other parameters relative to the reduced, unidentified, model. For ease of comparison we preserve, as before, $N = 100$ and $N_{Male}= 40$.
%----------------------------------------------------------------------------------------------------
\subsection{Application to tiger camera trapping data from Nagarahole} \label{NHda}
\subsubsection{Sampling design}
We have considered a specific application of modelling the bilateral capture-recapture data from a single season camera trapping study on tigers in Nagarahole national park of southern India (area = 1134 sq km). 
%WE USED XXXXX TYPE OF CAMERA TRAPS IN OUR STUDY. 
% Total area of the park  = 1134 sq km.
% Habitable area (state space) = 1127 sq km.
%(WHY ARE THESE THINGS IN CAPS?)\\
%(\textsc{I wanted to highlight these additions to Arjun. I have removed the caps now.})
The study area extends from 596626.7m to 641533.9m longitudinally and 1301307.5m to 1371205.7m latitudinally. The coordinates are in Universal Transverse Mercator (UTM) unit system. 
% $[596626.7, 641533.9]$ (East) x $[1301307.5, 1371205.7]$ (North) UTM units\\
% $[75.887, 76.302]$ (East) (longitude) x $[11.770, 12.401]$ (North) (latitude)\\
% $[75 ^{\circ} 53'13.2", 76 ^{\circ}18'07.2"]$ (East) (longitude) x $[11 ^{\circ}46'12.0", 12 ^{\circ}24'03.6"]$ (North) (latitude)
% (Figure~\ref{NHpark2})
The trapping array (Web Figure~12) consisted of 162 dual camera stations (where two opposite cameras are installed facing each other in each trap station) with a mean spacing of 1.5 km and the survey lasted 50 days (November 26, 2014 to January 13, 2015), resulting in 7364 trap nights of effort. We used Panthera branded passive motion sensor cameras (Model: V4, Maker: Panthera) in our study.

Our use of the Gaussian function implies that the buffer around the trapping array should, theoretically, be set at infinity. However, for practical reasons, this is usually set large enough so that individuals have a near zero probability of being exposed to the trapping array beyond such a buffer \citep{royle2009bayesian}. Accordingly, we set a buffer of 10 km (aiming for a width $>3\sigma$) around the trapping array for analysing the tiger data. Surprisingly, practitioners often misunderstand the reason for deciding on a buffer width. For example, in a recent camera trap SECR study of tigers \citep{lingaraja2017evaluating} in the same landscape, but at a different site, the authors set an arbitrarily small buffer (buffer width $<0.75\hat{\sigma}$), which is both statistically and ecologically indefensible.

Tigers can be individually identified by matching the unique patterns of flanks on both left and right sides. Researchers use software \citep{hiby2009tiger}) to assist in matching flank patterns from photographs and consequently obtain individual specific detection histories in standard spatial capture-recapture format \citep{royle2009bayesian}. 
% \citealt{royle2014spatial}\citealt{royle2009bayesian}) \citealt{gopalaswamy2012program}). 
However, since flank patterns are not identical on both sides of a tiger, at least one simultaneous detection of both side flanks over the course of camera trapping survey is needed to identify a tiger. A ``simultaneous detection'' is defined for an individual when the event time recorded by passive motion sensor cameras matches exactly (to the minute) for either flanks of an individual.
% EVENT TIME recorded BY XXX MATCHES EXACTLY (TO THE MINUTE) FOR EITHER FLANKS OF AN INDIVIDUAL. 
 Data were arranged in the format described by the sampling structure defined in Table \ref{sampledata}. 
\subsubsection{Analysis}
\paragraph{Analytical model} We used the covariate information on genders for the detected individuals. As male and female tigers do not share the same $\sigma$, i.e., do not have the same home-range size, we modelled $\sigma$ as a function of this covariate. We fit the model described in Section~\ref{gpid} and augmented the detection histories by all-zero detections to make them of same dimension. We ran one chain of 50000 iterations and discarded first 25000 as burn-in. 
% \ref{joint_posterior}
\paragraph{Inference}
%As recommended by \cite{chandler2013spatially},
We assessed inference from the parameter estimations by conducting `back simulations'. Here, we fixed the parameters at the estimated values and simulated 100 data sets under the same conditions. We computed coverage probabilities to assess the quality of the parameter estimates. Coverage probabilities are computed as the proportion of times when the estimated 95\% credible intervals contain the true value of the parameter. In these simulations, the true values are defined as the posterior mean estimates from the results of the field experiment of the following parameters: $N$, $\psi$, $N_{Male}$, $\theta$, $\phi$, $p_0$, $\sigma_m$, $\sigma_f$.% 
%--------------------------------------------------------------
\section{Results and Conclusions}\label{results}
\subsection{Assessment of model performance}\label{simresults}
\subsubsection{Simulation Results}
Here we brief the main findings of the simulation study over different simulation scenarios as mentioned in Section~\ref{simstudy}. The detailed discussion of the study is provided in Web Appendix C and the simulation results are presented in Web Table~3-10. We observe that, the quality of the estimates of different parameters significantly improves when trap entry probability $p_0$ increases. The scenarios in which $p_0$ is set to values greater than 0.03 had performed reasonably well. This is noted by the manner in which root mean square error (RMSE) values shrink significantly as $p_0$ increases. Whereas when the trap entry probability $p_0$ is set at low values (below 0.01), in most of those scenarios the posterior estimates of parameters are inaccurate with wide $95\%$ credible intervals. This outcome may be explained by the poor information content emerging when individuals rarely enter trap stations. The boxplots (Web Figure~2) of $N$, obtained by using the MCMC samples, show signs of positive skewness in most of the scenarios. Also, the bias and posterior standard deviation (SD) of $N$ are influenced by the conditional detection probability $\phi$ (indicating detector performance) in a similar manner to how $p_0$ influences model performance. That is, both bias and posterior SD decrease as the value of $\phi$ increases. 
%We define \emph{low} and \emph{high} values for $p_0 = \{0.005, 0.01\}$ and $\{0.05, 0.07\}$ respectively and  \emph{low} and \emph{high} values for $\phi= \{0.3, 0.4\}$ and $\{0.7, 0.8, 0.9\}$ respectively. Under such a classification, two cases are of interest, viz., when (a) $p_0$ is low but $\phi$ is high, (b) $p_0$ is high but $\phi$ is low. In case (a), posterior distribution of $N$ has a moderate RMSE estimate ($\approx 63.61$), whereas in case (b), although $\phi$ is low which decreases the detection rate, high value of $p_0$ has clearly more influence on the underlying model as the RMSE  estimates ($\approx 11.18$) are smaller than scenarios under (a). Scenarios with both $p_0$ and $\phi$ taking high values performed the best with RMSE estimate ($\approx 10.05$). This result perhaps highlights the relative importance of setting trap stations at optimal locations as compared with the choice of detector quality as far as the quality of estimates is concerned. 

The scenarios with $\sigma_m$ = 0.4 and $\sigma_f$ = 0.2 performed better than scenarios with $\sigma_m$ = 0.3 and $\sigma_f$ = 0.15 while estimating $N$, in the sense of having lesser RMSE estimate ($\approx 51.46$ for the former setting as compared to $\approx 24.86$ for the latter setting). This is perhaps associated with the fact that with less movement, both the number of detections and the number of distance classes recorded in data decrease. We would envisage the trap station layout also plays an important role in this assessment \citep{sun2014trap}.
% (CITE SUN ET AL 2014). 
%Apart from $N$, influence of $p_0$ is similar on $\theta$. The RMSE estimates of $\theta$ are $\approx 0.173$ for small values of $p_0$. The estimates show improvement for high values of $p_0$ with smaller RMSE estimate ($\approx 0.067$). Both of $\sigma_m$ and $\sigma_f$ have large RMSE estimates ($\approx 0.111$, $\approx 0.105$, respectively) for low values of $p_0$. Whereas for high values of $p_0$, the RMSE estimates are approximately $0.011$ and $0.015$, respectively. 
The estimated posterior correlations between $\phi$ and $p_0$ lie between $-0.3$ and $0$ for scenarios where both $p_0$ and $\phi$ take high values, i.e., $p_0 \in \{0.05, 0.07\}$ and $\phi \in \{0.7, 0.8, 0.9\}$. In comparison, for scenarios where $\phi$ takes small values, the posterior correlation estimates are between $-0.7$ and $-0.5$, whereas the RMSE estimates of $N$ are $\approx 11.18$. Here posterior mean estimates of $N$ show signs of robustness, even though a moderate amount of covariation is present between MCMC samples of $\phi$ and $p_0$.
%This shows that, whilst the estimates of $\phi$ and $p_0$ are being influenced when true values of $\phi$ is small, the estimates of $N$ still stays robust. 
%(THIS SENTENCE IS CONFUSING.)\\
%(\textsc{Thanks. I have rewritten the sentence.})

The posterior mean estimates of $p_0$ have a decreasing trend on bias as $\phi$ increases. In a similar manner, posterior mean estimates of $\phi$  also has a decreasing trend on bias as $p_0$ increases. This simulation outcome is indicative of poor information content in the data and consequently reflects on the identifiability of the parameter estimates. We surmise that these correlations will play an important role during model selection and inference. 
%---------------------------------------------------------------------------
\subsubsection{Comparison with `unidentified' model}
In both the scenarios, $\{\phi = 0.4,\, p_0 = 0.05, \, \sigma_m = 0.3, \sigma_f = 0.15\}$ and  $\{\phi = 0.3, \, p_0 = 0.05, \, \sigma_m = 0.4, \, \sigma_f = 0.2\}$, we see a substantial bias in the estimates of $N$ (see Table~\ref{compare2}) corresponding to the model which does not disentangle the parameters $p_0$ and $\phi$ \citep{royle2015spatial}. In these two scenarios the estimated posterior correlation between $\phi$ and $p_0$ are $-0.518$ and $-0.689$ respectively. Furthermore, we observe that the estimates of $\phi$ and $p_0$ are not unbiased, but the estimate of $N$ stays robust. The estimates of $\lambda_0$ in the unidentified model are $0.023$ and $ 0.016$ corresponding to the two scenarios which are close to the product of the true values of $\phi$ and $p_0$; also, the estimates of $N$ have larger RMSEs. These indicate that this model involves an over-simplification of the true model assumptions.

%Bilateral_Royle_MCMC_gender_150917v3_j4_sim27_ndraws30000_150917_113530
%sim	xlim.1.	xlim.2.	ylim.1.	ylim.2.	buffer	K	J	state.space.area	N.given	N.Male.given	numl	numr	IDfixed	known	phi.given	p0.given	sigmam.given	sigmaf.given	Msexsigma
%27	0	5	0	7	1	160	50	35	100	40	45	44	26	some	0.3	0.05	0.3	0.15	1
%Bilateral_Royle_MCMC_gender_150917v3_j9_sim26_ndraws30000_150917_113539
%sim	xlim.1.	xlim.2.	ylim.1.	ylim.2.	buffer	K	J	state.space.area	N.given	N.Male.given	numl	numr	IDfixed	known	phi.given	p0.given	sigmam.given	sigmaf.given	Msexsigma
%26	0	5	0	7	1	160	50	35	100	40	51	52	32	some	0.4	0.05	0.3	0.15	1
%Bilateral_Royle_MCMC_gender_150917v3_j39_sim2_ndraws30000_150917_113505
%sim	xlim.1.	xlim.2.	ylim.1.	ylim.2.	buffer	K	J	state.space.area	N.given	N.Male.given	numl	numr	IDfixed	known	phi.given	p0.given	sigmam.given	sigmaf.given	Msexsigma
%2	0	5	0	7	1	160	50	35	100	40	60	58	30	some	0.3	0.05	0.4	0.2	1
%Bilateral_Royle_MCMC_gender_150917v3_j44_sim26_ndraws30000_150917_113505
%sim	xlim.1.	xlim.2.	ylim.1.	ylim.2.	buffer	K	J	state.space.area	N.given	N.Male.given	numl	numr	IDfixed	known	phi.given	p0.given	sigmam.given	sigmaf.given	Msexsigma
%26	0	5	0	7	1	160	50	35	100	40	61	64	45	some	0.4	0.05	0.4	0.2	1
\begin{table}[!htbp] \centering 
  \caption{\textit{Posterior estimates for two different models are compared : (i) Unidentified model where detection probability parameter is defined as $p_{ik} = \lambda_0 \exp(-\frac{d(s_i, x_k)^2}{2\sigma^2})$, (ii) Identified model where trap entrance probability parameter is defined as $\pi_{ik} = p_0 \exp(-\frac{d(s_i, x_k)^2}{2\sigma^2})$, detection probability conditional on trap entrance is defined as $\phi$. Movement parameter is $\sigma_m$ or $\sigma_f$, depending on whether individual is male or female, respectively.)}}
%  \textit{Posterior estimates of parameters. Unidentified model: $p_{ik} = \lambda_0 \exp(-\frac{d(s_i, x_k)^2}{2\sigma^2})$, Our model: $\pi_{ik} = p_0 \exp(-\frac{d(s_i, x_k)^2}{2\sigma^2})$, $\sigma = \sigma_m$ or $\sigma_f$ depending on individual is male or female.} \\
%  (ALSO SAY THAT OUR $p_{ik} = \phi_k \pi_{ik}$?)\\
%  (\textsc{Is the following alright ?} \\
{\footnotesize \begin{tabular}{ l @{\extracolsep{20pt}} c @{\extracolsep{20pt}} c  @{\extracolsep{20pt}} c @{\extracolsep{20pt}} c} 
\\[-1.8ex]\hline 
\hline \\[-1.8ex] 
 & \multicolumn{2}{c}{(i) Unidentified model\hspace{50pt} }  & \multicolumn{2}{c}{(ii) Identified model\hspace{25pt} } \\ [1.2ex] 
 & \multicolumn{2}{c}{($p_0$ and $\phi$ not identified)\hspace{50pt} }  & \multicolumn{2}{c}{($p_0$ and $\phi$ identified)\hspace{25pt} } \\ [1.2ex] 
Parameters & Mean & RMSE & Mean &  RMSE \\ 
[1.2ex] \hline \\[-1.8ex] 
\multicolumn{5}{l}{Scenario : $\phi = 0.4$, $p_0 = 0.05$, $\sigma_m = 0.3$, $\sigma_f = 0.15$} \\ [1.5ex] 
$N$ & $89$ & $12.743$ & $99$ & $10.060$ \\ 
$N_{Male}$ & $36$ & $5.726$ & $40$ & $4.109$ \\ 
$\phi$ & - & - & $0.352$ & $0.059$ \\ 
$p_0$ & - & - & $0.060$ & $0.013$ \\ 
$\lambda_0$ & $0.023$ & $0.027$ & - & - \\ 
$\sigma_m$ & $0.287$ & $0.017$ & $0.307$ & $0.016$  \\ 
$\sigma_f$ & $0.157$ & $0.010$ & $0.144$ & $0.011$  \\ 
[1.2ex] \hline \\[-1.8ex] 
\multicolumn{5}{l}{Scenario : $\phi = 0.3$, $p_0 = 0.05$, $\sigma_m = 0.4$, $\sigma_f = 0.2$} \\ [1.5ex] 
$N$ & $93$ & $9.615$ & $99$ & $7.632$ \\ 
$N_{Male}$ & $43$ & $4.405$ & $40$ & $3.464$ \\ 
$\phi$ & - & - & $0.283$ & $0.035$ \\ 
$p_0$ & - & - & $0.047$ & $0.007$ \\ 
$\lambda_0$ & $0.016$ & $0.033$ & - & - \\ 
$\sigma_m$ & $0.364$ & $0.039$ & $0.407$ & $0.018$  \\ 
$\sigma_f$ & $0.199$ & $0.010$ & $0.216$ & $0.021$  \\ 
[1.2ex] \hline \\[-1.8ex] 
\end{tabular}}
\label{compare2} 
\end{table}
\subsection{Application to tiger camera trapping data from Nagarahole}\label{daresults}
\subsubsection{Data summary}
In our field experiment, we could identify 65 tigers (22 male, 33 female, 10 of unknown sex). This meant that we recorded at least one simultaneous capture of both flanks for each of the above set of tigers. In addition, we obtained 14 partially identified left flank only detection histories (6 male, 5 female, 3 of unknown sex) and 17 partially identified right flank detection histories (7 male, 4 female, 6 unknown). Overall, we obtained 123 simultaneous detections, 126 left flank only detections, and 137 right flank only detections.
\subsubsection{Data analysis}
The posterior estimates of parameters are provided in Table~\ref{NH15.post.stats}. The posterior mean estimate of population size (over the state space) is 133 with a 95\% credible interval of (117, 152). The density of tiger is estimated at 11.73 tigers per 100 km$^2$ in our study area. The posterior mean of $\sigma_m$ ($1.970$) is estimated to be higher than that of $\sigma_f$ ($1.209$). 
%We observe that male tigers had a movement parameter $\sigma$ that was 1.5 times larger than for females, which indicates heterogeneity in the extent of movements from their activity centres. 
The estimates of $\sigma_m$ and $\sigma_f$ also confirm that the buffer we had set (10 km) was sufficiently large enough. The number of male tigers in the population is estimated at 41 with a 95\% credible interval (33, 50), and hence the number of female tigers is estimated at 92.  The sex ratio was estimated to be 2.24 females to 1 male. 

The scatter plot provided in Web Figure~15 
%Figure~\ref{NH15.scatplot}
shows that there is moderate amount of correlation between $\phi$ and $p_0$ ($\approx -0.41$) present in the MCMC samples, which also matches the simulation results for relatively smaller values of $\phi$. Sample correlation between the pairs ($p_0$, $\sigma_m$) ($\approx -0.44$) and ($ p_0$, $\sigma_f$) ($\approx -0.48$) indicate identifiability issues between those parameters, but is not expected to effect the estimate of the other parameters of interest viz., $N$, $N_{Male}$. As we discussed in Section~\ref{identifiability:p0_sigma}, the accuracy and precision of $p_0$ and $\sigma$'s depend on dispersion of distances between individuals' activity centres and trap locations. Higher dispersion in these distances is likely to make the estimates of $p_0$, $\sigma_m$ and $\sigma_f$ more accurate and precise.
% (PROVIDE A STATEMENT ON THE COVARIATION BETWEEN OTHER PARAMETERS)
\subsubsection{Inference}
The detection probability $\phi$ in the analysis of Nagarahole capture-recapture data set on tigers is estimated at 0.489 (see Table~\ref{NH15.post.stats}). This implies that each camera records a clear flank image in a little less than 50\% of the cases. This is not surprising to us as a clear `valid sample' depends on many other factors, such as quality of the traps, camera malfunctions, ambient temperature etc. in typical field conditions.
 
The simulation study was designed to reflect a typical field study, so that performance of the model can be evaluated based on different values taken by the model parameters in a practical setup. Accordingly, $p_0$ is the most dominant parameter which influences the performance of the model while obtaining posterior summaries of the other parameters. Furthermore, the estimates corresponding to the scenarios where $p_0$ is set to 0.05 or 0.07 perform fairly well as compared to the scenarios where $p_0$ is set to smaller values, viz., 0.005, 0.01. In the field study $p_0$ is estimated at 0.041 with a 95\% credible interval (0.033, 0.049) (see Table~\ref{NH15.post.stats}).

We estimated tiger density to be 11.73 tigers per 100 km$^2$ in our study area. This is comparable to estimates of tiger density from other similar studies in this area  using other versions of SECR models (\citealt{royle2009bayesian}, \citealt{dorazio2017hierarchical}). We found that coverage probabilities of all the continuous parameters (viz., $\psi$, $\theta$, $\phi$, $\sigma_m$, $\sigma_f$), except $p_0$, attained the nominal coverage probability 0.95 (see Table~\ref{NH15.post.stats}) 
while coverage probabilities of $p_0$, $N$ and $N_{Male}$ are 0.91, 0.88 and 0.90, respectively, which do not attain the nominal level. This conservative coverage probabilities are indicative of imprecise credible intervals. However, coverage probabilities are expected to increase with a better detection and trap entry rates (i.e., higher $\phi$ and $p_0$) as we have discussed above. 
%da_codes/results_NW15_da_301116/lr.gender_NW15_301216_133035
\begin{table}[!htbp] \centering 
  \caption{\textit{Posterior estimates of parameters from the Nagarahole tiger analysis}} 
{\footnotesize \begin{tabular}{ l@{\extracolsep{5pt}}c c c c c c c} 
\\[-1.8ex]\hline 
\hline \\[-1.8ex] 
Parameters & Mean & SD & 2.5\% & 50\% & 97.5\% & CI width & Coverage probability  \\ 
\hline \\[-1.8ex] 
$N$ & $133$ & $8.89$ & $117$ & $133$ & $152$ & $35$ & $0.880$\\ 
$\psi$ & $0.333$ & $0.032$ & $0.272$ & $0.332$ &  $0.397$ & $0.125$ & $0.980$ \\ 
$N_{Male}$ & $41$ & $4.293$ & $33$ & $41$ & $50$ & $17$ & $0.900$\\ 
$\theta$ & $0.312$ & $0.050$ & $0.220$ & $0.310$ & $0.413$ & $0.193$ & $1$ \\ 
$\phi$ & $0.486$ & $0.029$ & $0.430$ & $0.487$ & $0.543$ & $0.113$ & $0.950$ \\ 
$p_0$ & $0.041$ & $0.004$ & $0.033$ &  $0.040$ & $0.049$ & $0.015$ & $0.910$ \\ 
$\sigma_m$ & $1.970$ & $0.083$ & $1.814$ & $1.967$ & $2.140$ & $0.326$ & $0.960$ \\ 
$\sigma_f$ & $1.209$ & $0.056$ & $1.103$ & $1.207$ & $1.323$ & $0.220$ & $0.940$ \\ 
\hline \\[-1.8ex] 
\end{tabular}}
\label{NH15.post.stats} 
\end{table} 
\begin{figure}[!htbp]
\centering
\begin{tabular}{l}
\includegraphics[width=450pt,height=450pt]{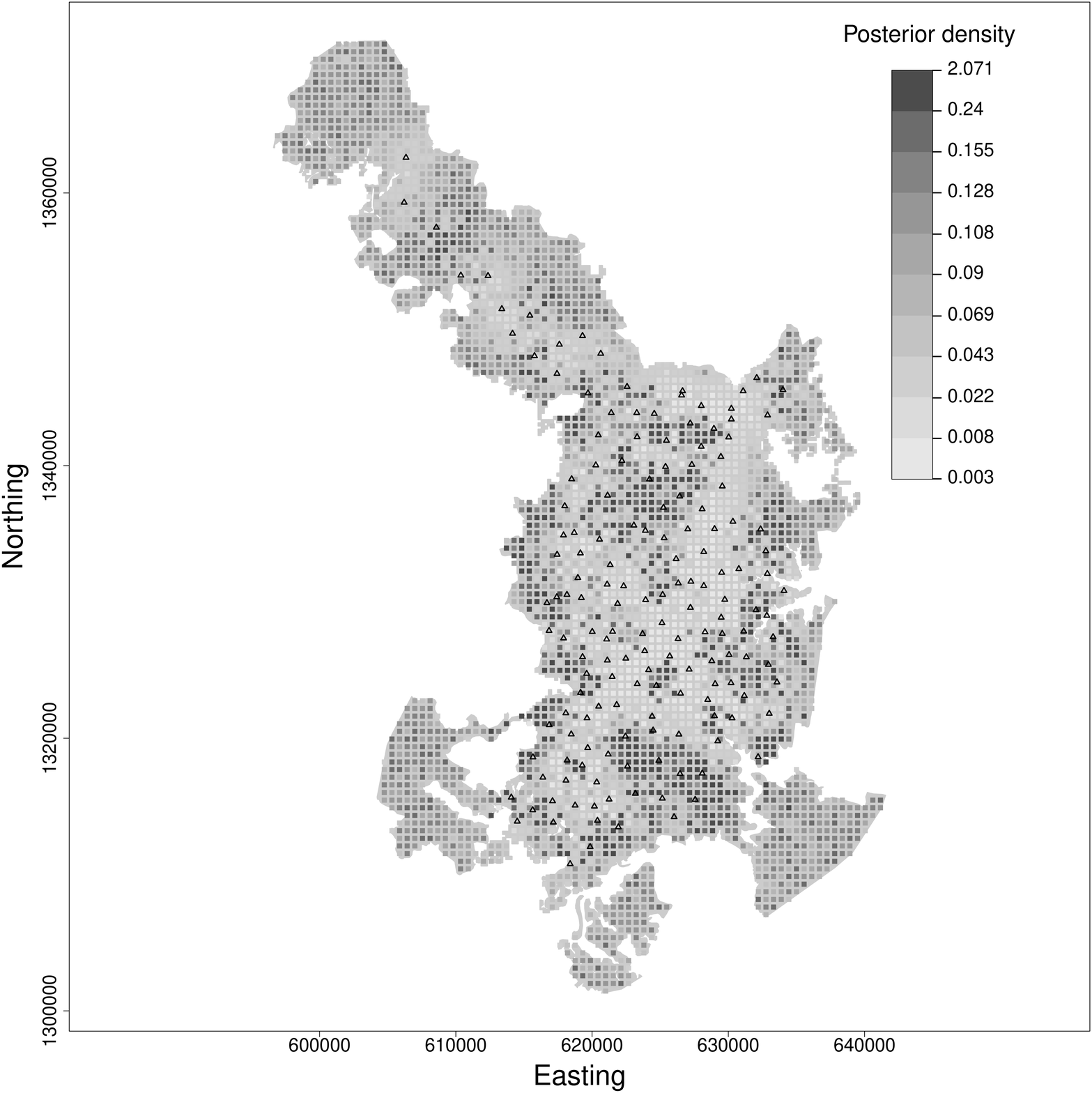}
\end{tabular}
	\caption{Estimated posterior density of tigers (per km$^2$) in Nagarahole reserve. Each pixel is marked with $E [N(s)|\text{data}]$ where $N(s)$ is the number of activity centres situated at pixel $s$. Lighter grey shade (white) indicate lower densities, while darker grey shade (black) correspond to higher densities. The trap stations are denoted by `\small$\triangle$'.}
	\label{NHpark_density_mapp_grey80}
\end{figure}
\section{Discussion} \label{discussion}
In this article, we have developed a novel SECR estimator that successfully disentangles the ecological process of animal trap entry from the observation process of trap detection rates. 
%(CITE ROYLE AND DORAZIO 2008, see METHODS).
Our simulation results highlight the relative importance of ensuring that trap stations are chosen based on good locations as compared to the importance of detector choice, especially, when there is more than one detector located at each station. 
Our SECR model is built upon an earlier Bayesian hierarchical model by \cite{royle2015spatial} and makes full use of all data available (including information on partially identified individuals). We demonstrate how our model provides unbiased estimates of population size $N$ when trap detection rate is less than one. We justify the importance of estimating trap detection rate $\phi$ by showing the bias in the estimate of $N$ when we use the \cite{royle2015spatial} model under certain simulation conditions. 

We have developed the estimator using the special case of having only two detectors at each station, each detector capturing a set of unique traits about the identities of individuals. The assumption, however, is that each detector contains enough information on its own to ascertain individual identity. For example, as this study was motivated by the tiger example we have discussed in the paper, we find a field situation where two profile flanks of an individual tiger are attempted to be caught at the same time at trap stations. When we do not have simultaneous captures it is not possible to tell if a right flank image of a tiger has an equivalent left flank image or not. 
%However, we can tell \emph{without any doubt} if two right flank images belong to the same tiger or not. 
We recognize that the situation will not directly apply if the same idea is extended to genotyping problems (\citealt{mondol2009evaluation}, \citealt{Sethi2016genetic}) because at each locus there is not enough information to convincingly identify individuals. We discuss more on this application later. 

It is possible to extend our model to include three or more detectors per station based on the idea of how occupancy models \citep{mackenzie2002estimating} were constructed to include multiple sampling occasions. However, we envisage some complications with regard to explicitly defining the permutative arrangement of capture histories. For this, we need to understand how many detectors (implying how many sets of unique features) are necessary to establish full identity of an individual. For example, in genotyping problems \citep{Sethi2016genetic}, workers identify a panel of loci to achieve a desirably low level of probability of identity ($\mathrm{P}_{ID}$). During field surveys \citep{mondol2009evaluation},
% \citealt{gardner2010spatially}),
workers often gather faecal samples for subsequent genotyping. However, not all faecal samples amplify in the laboratory. We envisage the application of our model to estimate this probability using the parameter $\phi$. 

As with most estimators, the utility of our SECR model is enhanced when meaningful covariates are applied on the specific model parameters. Ecologists interested in obtaining an understanding about fine scale movements of animals can now do so without the worry about the confounding problem of detector efficiency. We envisage that our estimator will find much use in optimal allocation problems \citep{augustine2016spatial} in wildlife surveys. For example, many camera traps are available in the market at various costs. Since our model specifically estimates a parameter $\phi$ associated with trap efficiency, it would come of use to evaluate the relative gains in precision of estimates of abundance when, for example, cheap cameras are replaced by expensive cameras or to decide how many traps are needed at each station. Further, for defined monitoring budgets our model can be used to determine the most optimal allocation of the number of trap stations and the types of traps with available resources.
%for the problems on hand. 

Beyond ecology, our SECR estimator lays the foundation for solving the statistical reconciliation problem in administrative lists \citep{york1992bayesian}. 
%(CITE YORK AND MADIGAN 1992).
In this problem, individuals do appear in different administrative lists at a region and the problem is to identify the population size from captures of individuals in the multiple lists. We find equivalence between multiple detectors discussed in our problem with the presence of different administrative lists in the problem described in \cite{york1992bayesian}.
% YORK AND MADIGAN 1992.
%In addition, the spatially explicit treatment of our problem enables an additional dimension previously not accounted for in such problems.  We also envisage the application of this estimator in disease problems  (\citealt{mcclintock2010seeking}, \citealt{banerjee2014hierarchical}),
% (eg: CITE MCLINTOCK ET AL 2010, BANNERJEE BOOK),
%where there are hierarchical layers present during detection of disease in individuals over a region. 

An inherent problem in the application of a complex model for real world problems, and a larger problem in the statistical literature, is that selecting the appropriate model for prediction and characterization of populations is not straightforward. Some of us are currently working on evaluating and applying various model selection tools on this class of Bayesian SECR problems. We also encourage the extension of this estimator to include multiple detectors (more than two) as described above. With these developments, we envision wide application of the general approach presented here. %
\section{Acknowledgements}
We thank the Indian Statistical Institute for financial and administrative support and the Centre for Wildlife Studies and Wildlife Conservation Society, India program for providing the data and analytical support. We also thank Ravishankar Parameshwaran and Devcharan Jathanna from Centre for Wildlife Studies for help with the computer simulations and helpful advice. AMG thanks the Wildlife Conservation Society, New York for partial funding support.
\section{Supplementary Materials}
Web Appendices, Tables, and Figures, referenced in Sections~\ref{methods} and \ref{results}, and the implemented R codes
% implementing the new methods
  are available with this paper at the Biometrics website on Wiley Online Library.
\bibliographystyle{biom}
\bibliography{bilateral_bibliography} %.bib}
\label{lastpage}
 \end{document}